\documentclass[aps, prl, twocolumn, superscriptaddress, longbibliography]{revtex4-1}
\usepackage[colorlinks, linkcolor=blue, anchorcolor=blue, citecolor=blue]{hyperref}
\usepackage{amsmath}
\usepackage{graphicx}
\usepackage{braket}
\usepackage{multirow}
\usepackage{color}
\usepackage{soul}

\begin{document}

\title{Hund's metal physics: from SrNiO$_2$ to NdNiO$_2$}
\author{Y. Wang}     
\affiliation{Department of Condensed Matter Physics and Materials Science, Brookhaven National Laboratory, Upton, New York 11973, USA} 

\author{C.-J. Kang}
\affiliation{Department of Physics and Astronomy, Rutgers University, Piscataway, New Jersey 08856, USA}

\author{H. Miao}
\affiliation{Material Science and Technology Division, Oak Ridge National Laboratory, Oak Ridge, Tennessee 37830, USA} 

\author{G. Kotliar}
\affiliation{Department of Condensed Matter Physics and Materials Science, Brookhaven National Laboratory, Upton, New York 11973, USA} 
\affiliation{Department of Physics and Astronomy, Rutgers University, Piscataway, New Jersey 08856, USA}

\date{\today}

\begin{abstract}
We study the normal state electronic structure of the recently discovered infinite-layer nickelate superconductor, Nd$_{1-x}$Sr$_x$NiO$_2$. Using SrNiO$_2$ as a reference, we find that Nd$_{1-x}$Sr$_x$NiO$_2$ is a multi-orbital electronic system with characteristic Hund's metal behaviors, such as metallicity, the importance of high-spin configurations, tendency towards orbital differentiation, and the absence of magnetism in regimes which are ordered according to static mean-field theories. In addition, our DFT+DMFT calculations with exact double counting scheme show that despite large charge carrier doping from SrNiO$_2$ to LaNiO$_2$, the Ni-$3d$ total occupancy is barely changed due to the decreased hybridization with the occupied oxygen-2$p$ states, and increased hybridization with the unoccupied La-5$d$ states. Our results are in good agreement with the existing resonant inelastic x-ray scattering measurements and pave the way to understand the pairing mechanism of Nd$_{1-x}$Sr$_x$NiO$_2$.
\end{abstract}

\maketitle
\textit{Introduction}---Understanding high temperature (high-$T_{c}$) superconductors is an outstanding question in the area of quantum materials and presents major unsolved questions. With the discovery of new materials, multiple questions arise: what are the origins of the high-$T_{c}$ superconductivity and what are the competing orders. Understanding of these questions requires a proper description of the low-energy electronic structure and the correlations presented in the normal state. The recent discovery of superconductivity in the infinite-layer nickelates, Nd$_{1-x}$Sr$_x$NiO$_2$, by Hwang's group~\cite{Li2019} raises these questions in a novel context~\cite{Norman_trend:2020}. 

We have by now, at least four broad classes of unconventional superconducting materials, which have  well established phenomenologies: (1) the quasi-one dimensional  organics \cite{Jerome:1991}; (2) materials proximate to a Mott-Hubbard or a charge-transfer insulator~\cite{Zaanen:1985}  such as the quasi-two dimensional organic salts~\cite{Powell:2006} and the copper oxide based materials~\cite{Bednorz:1986,Patrick:2006,Dagotto:1994} which can be described, at low energies, in terms of an effective single band system; (3) intermetallics including rare-earth or actinides heavy-fermion systems~\cite{stewart:1984}; and (4) the more recently discovered iron pnictides and chalcogenides~\cite{Kamihara:2008,stewart:2011}. Numerous efforts have been made to synthesize  $3d^9$ nickelate materials that can serve as analogs of the doped copper oxide superconductors, such as La$_4$Ni$_3$O$_8$~\cite{Poltavets:2010,Junjie:2019}.

Early on, LaNiO$_2$ was suggested theoretically to be a cuprate analog by Ansisimov \textit{et al.}~\cite{Anisimov:1997} .  However, Lee and Pickett~\cite{Pickett:2004} showed that the electronic structure of LaNiO$_2$, differs significantly from the cuprates, as rare-earth $5d$ bands were shown to cross the Fermi level as well. Other differences include smaller $c$-axis and more three-dimensional nature, smaller crystal field splitting and much smaller hybridization strength between the Ni-3$d$ and oxygen-$2p$ states~\cite{Botana:2020}. Alternatively, analogies with heavy-fermion materials have also been proposed~\cite{GuangMing:2020}. Despite extensive investigations by experiments~\cite{Hepting2020,YingFu:2019,ZengSW2020,Dangfeng:2020,Osada:2020,Berit:2020,Kyuho:2020,LiQing:2020,Bixia:2020}, density functional theory (DFT)~\cite{Peiheng:2019,Jiacheng:2019,Nomura:2019,LiuZhao:2020,Xianxin:2020,MiYong:2020,MiYongb:2020,Botana:2020,Hirayama:2020,Emily:2020,Benjamin:2020}, DFT plus dynamical mean-field theory (DFT+DMFT)~\cite{georges:1996,kotliar:2006,lichtenstein:2001} methods~\cite{Karp:2020,SiLiang:2020,Ryee:2020,GuYuhao:2020,Leonov:2020,Leonov_b:2020,Motoharu:2020,Lecherman:2020a,Lecherman:2020} and theoretical models~\cite{LunHui:2019,Hirofumi:2019,Jiangmi:2019,Werner:2020,ZhangHu:2020,ZhangYahui:2020,GuangMing:2020}, the nature of the electronic correlations in Nd$_{1-x}$Sr$_x$NiO$_2$ still remains an open question. In particular, a quantitative agreement between experiment and theory is still absent.

In this work, we carry out various electronic structure calculations and analyze the available data for the infinite-layer nickelates at the two ends of hole-doping (LaNiO$_2$ and SrNiO$_2$, respectively) as well as the intermediate doping regime. We point out  multiple similarities between (La,Sr)NiO$_2$ and the Hund's metal iron-based superconductors~\cite{Hoshino:2015,Werner:2020,Tsunghan:2018}, once we take into account obvious differences,  the former has a valence close to $3d^{8.5}-3d^9$ and with low-energy $e_g$ orbitals while the later has Fe $d^{5.5}-d^{6.5}$ valence with low-energy $t_{2g}$ orbitals~\cite{Kamihara:2008,stewart:2011,Qimiao:2016}.

\begin{figure*}
\centering
\includegraphics[width=0.98\textwidth]{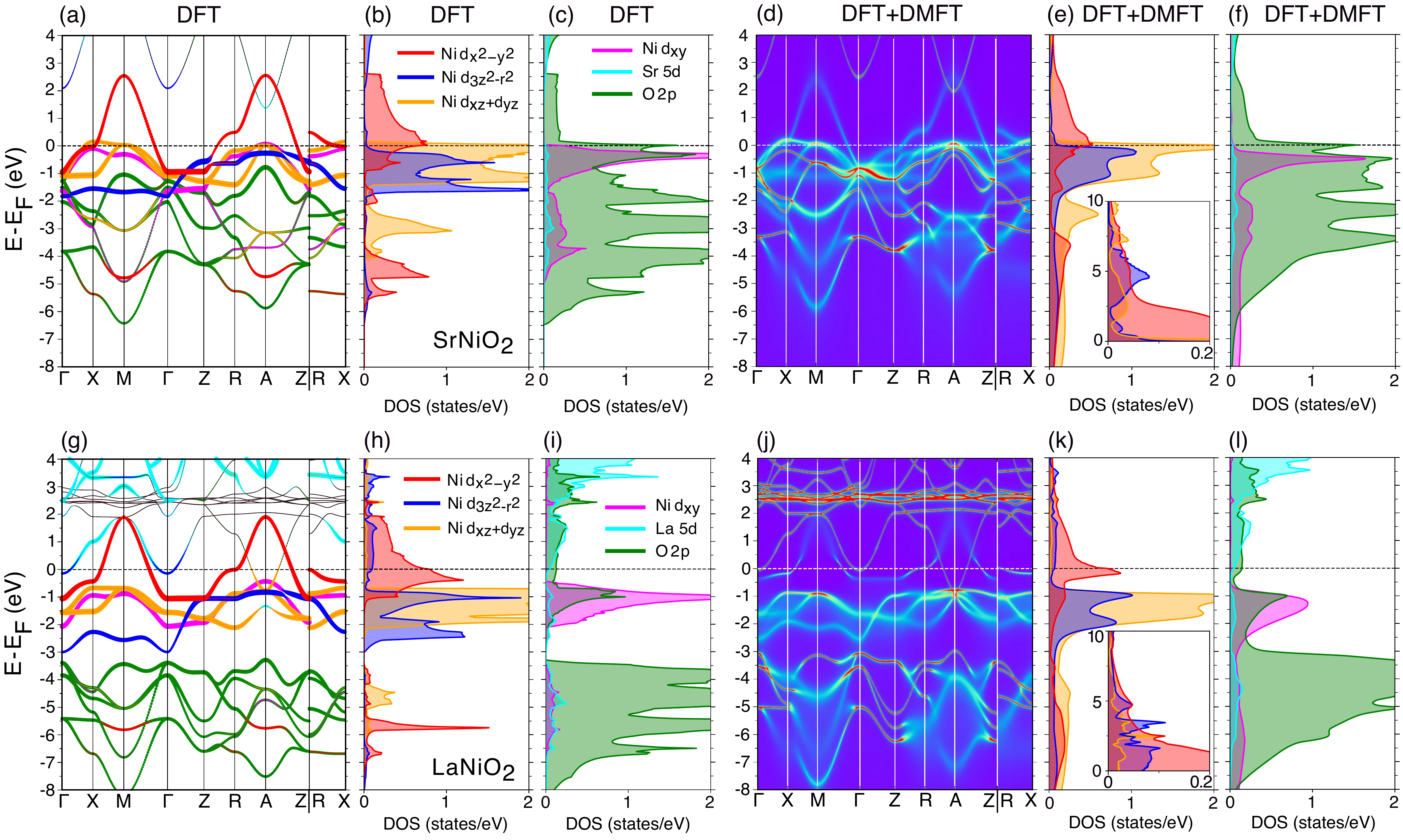}
\caption{DFT and DFT+DMFT band structures/$k$-resolved spectrum functions and density of states. Upper (lower) panels for SrNiO$_2$ (LaNiO$_2$). (a),(g) DFT band structures and (b),(c),(h),(i) DFT density of states. (d),(j) DFT+DMFT $k$-resolved spectrum functions and (e),(f),(k),(l) DFT+DMFT density of states. The insets in (e) and (k) are the enlarged density of states above the Fermi level from 0 to 10 eV.}
\label{fig:band}
\end{figure*}

\textit{Methods}---We study LaNiO$_2$ instead of NdNiO$_2$ to avoid the issue of Nd-$4f$ orbitals~\cite{Botana:2020, Ryee:2020}. Although SrNiO$_2$ does not crystallize in the LaNiO$_2$ structure~\cite{Pausch:1976}, in this work, for the purpose of comparison, we assume the same crystal structures ($P4/mmm$, see Fig.S1 in~\cite{suppl}) and use the same lattice parameters $a=b=3.966\text{\AA}$ and $c=3.376\text{\AA}$~\cite{Levitz:1983,Botana:2020} for both materials. We perform fully charge self-consistent DFT+DMFT calculations using the code, EDMFTF, developed by Haule \textit{et al.}~\cite{Haule:2010} based on the Wien2k package~\cite{Blaha:2020}. We choose a wide hybridization energy window from -10 eV to 10 eV with respect to the Fermi level  to cover the long tails of Ni $d_{3z^2-r^2}$ and $d_{xz/yz}$ orbitals due to their strong hybridizations with La/Sr $5d$ states and interstitial-$s$ states~\cite{GuYuhao:2020}. All the five Ni-$3d$ orbitals are considered as correlated ones and a local Coulomb interaction Hamiltonian with rotationally invariant form is applied. We choose $U=5$ eV and Hund's coupling $J_{H}=1$ eV, which are reasonable values for this system~\cite{Hirofumi:2019,Leonov:2020}. The local Anderson impurity model is solved by the continuous time quantum Monte Carlo (CTQMC) solver~\cite{Gull:2011, Haule:2010}. We use an exact double counting (DC) scheme invented by Haule~\cite{Haule:2015}, which nearly eliminates the DC issues in most correlated materials. The DFT+DMFT results shown in this work are performed at the temperature $T=100$ K. The self-energy on real frequency $\Sigma(\omega)$ is obtained by the analytical continuation method of maximum entropy~\cite{Haule:2010}. The local effective mass enhancement due to electronic correlation effects is directly obtained from the self-energy on Matsubara frequency, $m^*/m^{\text{DFT}}=1/Z=1-\frac{\partial{\text{Im}\Sigma(i\omega_n)}}{\partial{\omega_n}}|_{\omega_n\rightarrow0}$, to avoid the large error bar in analytical continuations. The Sr-doping effects are simulated by the virtual crystal approximation (VCA) in the intermediate doping regime ($x\le0.5$).
 
We also perform spin-polarized calculations to check the energetics of long-range magnetic orders on the static mean-field level using the GGA+$U$ method implemented in the VASP package~\cite{kresse:1996}. A $\sqrt{2}\times \sqrt{2}\times 2$ supercell is used~\cite{Ryee:2020, Botana:2020}. Four magnetic configurations (see Fig.S2 in~\cite{suppl}) are considered: (i) ferromagnetic (FM), (ii) FM in plane and antiferromagnetic out of plane (AFM-A), (iii) AFM in plane and FM out of plane (AFM-C) and (iv) AFM in plane and AFM out of plane (AFM-G). 

\textit{Multi-orbital nature and Hund's metal behavior}---We start with describing the electronic structure and correlation effects of the material at the extremely hole-doping end, SrNiO$_2$, as a reference system. Figs.~\ref{fig:band}(a)-(c) show its DFT band structure and density of states (DOS) that are similar to those of CaCuO$_2$~\cite{Botana:2020}, but with one electron removed (nominal $3d^8$). It shows an obvious multi-orbital electronic structure with the Fermi level crossing both the $e_g$ and $t_{2g}$ orbitals. The hybridization between Ni-$3d$ and Sr-$5d$ states is very weak. While, the Ni-$3d$ states strongly hybridize with the O-$2p$ states with a sharp peak in the O-$2p$ DOS at the Fermi level [see Fig.~\ref{fig:band}(c)], such that the charge-transfer energy, $\Delta_{dp}$, between the Ni-$3d$ and O-$2p$ states is small ($2\sim3$ eV)~\cite{Botana:2020} compared to Hubbard $U$, which indicates a possible charge-transfer Mott-insulator scenario like that in NiO~\cite{Imada:1998,Schuler:2005} if a large Mott gap could be opened. However, our DFT+DMFT calculations show that it is a moderately correlated multi-orbital metal. 

Table~\ref{tab:mass} show the local occupancy numbers, $n_d$, and the mass enhancement, $m^*/m^{\text{DFT}}$, of Ni-$3d$ orbitals, and Table~\ref{tab:multi} show the probability of the Ni-$3d$ local multiplets, obtained from the DFT+DMFT calculations. The strong $3d$-$2p$ hybridization leads to a much larger Ni-$3d$ total occupancy (8.479) than the nominal value. Except for the $d_{xy}$ orbital, all the other $3d$ orbitals are partially occupied and moderately renormalized due to the comparable strengths of Hund's coupling and crystal field splitting. There are 1.057 (0.943), 1.755 (0.245) and 1.845 (0.155) electrons (holes) residing on the $d_{x^2-y^2}$, $d_{3z^2-r^2}$ and $d_{xz/yz}$ orbitals, respectively. These holes are highlighted by the DOS above the Fermi level in the inset of Fig.~\ref{fig:band}(e). In the presence of large Hund's coupling, they form substantial spin-triplet states ($S=1$) with a weight of 25.4\%, while smaller weight (17.1\%) of spin-singlet states in the $N=8$ sector (see Table~\ref{tab:multi}). The Hund's effect is also be seen in the DFT+DMFT spectrum functions [see Figs.~\ref{fig:band}(d) and~\ref{fig:band}(e)], where the $d_{3z^2-r^2}$ band is pushed  closer to the Fermi level compared to the DFT results [Figs.~\ref{fig:band}(a)].  We find strong charge fluctuations with 7.7\%, 43.7\% and 5.6\% multiplets in the $N=7$, 9 and 10 sectors, respectively. The electronic correlation strengths exhibit  orbital differentiation, with a larger mass enhancement of 1.89 for $d_{x^2-y^2}$ orbital and a smaller one of about 1.6 for $d_{3z^2-r^2}$ and $d_{xz/yz}$ orbitals. Fig.~\ref{fig:band}(d)-(f) show that all the $3d$ orbitals except $d_{xy}$ contribute to the DFT+DMFT Fermi surfaces (see Fig.S3 in~\cite{suppl}) and that $\Delta_{dp}$ is further reduced by the electronic correlation effects. 

Thus, SrNiO$_2$ has many things in common with a correlated multi-orbital metal and  manifests Hund's metal behavior~\cite{werner:2008,haule:2009,medici:2011,yin:2011,georges:2013,Medici:2017,stadler:2019}, analogous to iron pnictides and chalcogenides, since it (i) is away from half-filling; (ii) has significant enhancement of electronic correlation and mass; (iii) shows important roles of Hund's coupling and high-spin configurations; (iv) shows orbital differentiation behavior.

We now go to the other end of the undoped LaNiO$_2$ and examine the evolution of band structures from SrNiO$_2$ to LaNiO$_2$ by comparing the DFT band structure and DOS in Figs.~\ref{fig:band}(g)-(i) to those in Figs.~\ref{fig:band}(a)-(c). The main changes are (i) the Fermi level moves up; (ii) the Ni $d_{x^2-y^2}$ band is pushed away slightly  from the other $3d$ bands; (iii) the La-$5d$ bands start to strongly hybridize with the Ni-$3d$ bands and contribute to the Fermi surfaces [see $\Gamma$ and A points in Fig.~\ref{fig:band}(g)]; (iv) the O-$2p$ bands are pushed down a lot to be nearly isolated from the Ni-$3d$ bands and the $3d-2p$ hybridization becomes much weaker such that $\Delta_{dp}$ is significantly reduced. In this work, we also reveal a similar multi-orbital nature of SrNiO$_2$, even in the undoped LaNiO$_2$.

\begin{table}
    \centering
    \caption{The local occupancy numbers, $n_d$, and the mass enhancement, $m^*/m^{\text{DFT}}=1/Z$, of Ni-$3d$ orbitals from DFT+DMFT calculations.}
 \setlength{\tabcolsep}{0.74mm}{\begin{tabular}{c|ccccccc}
        \hline
        \hline
                     &   & $d_{x^2-y^2}$ & $d_{3z^2-r^2}$ & $d_{xz}$ & $d_{yz}$ & $d_{xy}$ & Total \\
         \hline
         \multirow{2}{*}{$n_{d}$}  & SrNiO$_2$ & 1.057 & 1.755 & 1.845 & 1.845 & 1.977 & 8.479 \\ 
                 & LaNiO$_2$ & 1.219 & 1.637 & 1.891 & 1.891 & 1.955 & 8.593 \\
         \hline
         \multirow{2}{*}{$m^{*}/m^{\text{DFT}}$} &SrNiO$_2$& 1.89 & 1.56 & 1.58 & 1.58 & 1.32 & \\  
                                &LaNiO$_2$& 2.81 & 1.25 & 1.21 & 1.21 & 1.27 & \\
         \hline
         \hline
    \end{tabular}}
    \label{tab:mass}
\end{table}

\begin{table}
\centering
\caption{The weights of the Ni-$3d$ local multiplets sampled by CTQMC solver.}
\setlength{\tabcolsep}{1.9mm}{\begin{tabular}{cccccc}
\hline 
\hline
Occupancy $N$ & $7$ & $8$ & $8$ & $9$ & $10$\\
Total Spin $S$ & All & $1$ & $0$ & $1/2$ & $0$ \\
\hline 
SrNiO$_2$, DMFT & $7.7\%$ & $25.4\%$ & $17.1\%$ & $43.7\%$ & $5.6\%$\\
\hline 
LaNiO$_2$, DMFT & $6.2\%$ & $25.9\%$ & $10.3\%$ & $49.0\%$ & $8.1\%$\\
LaNiO$_2$, Ref.~\cite{Hepting2020} & $6.0\%$ & $24.0\%$ & $14.0\%$ & $56.0\%$ & \textendash{}\\
\hline 
\hline
\end{tabular}}
\label{tab:multi}
\end{table}

\begin{figure}
\centering
\includegraphics[width=0.48\textwidth]{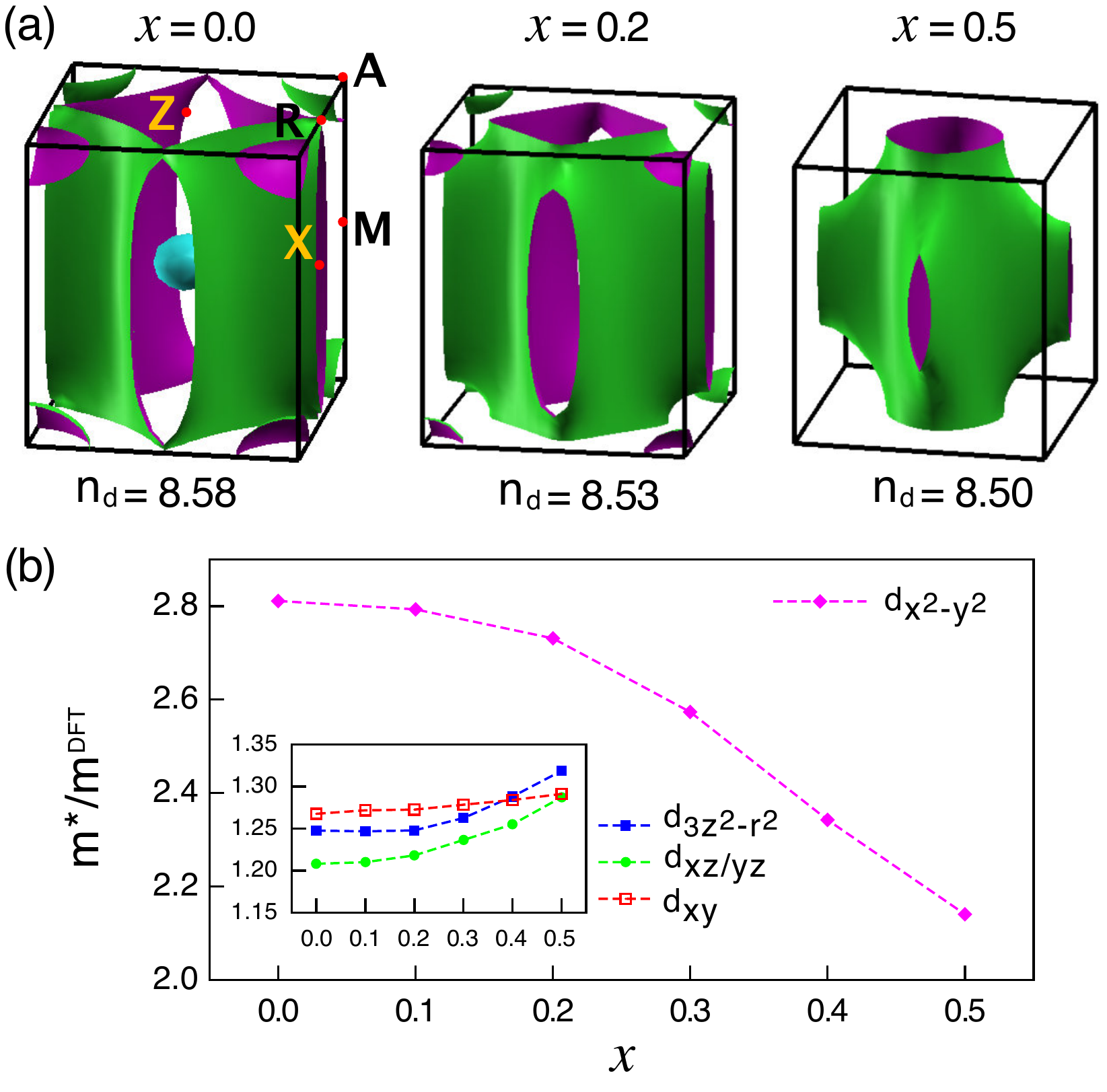}
\caption{(a) Evolution of the DFT+DMFT Fermi surfaces and (b) mass enhancement of Ni-$3d$ orbitals as functions of doping level $x$.}
\label{fig:doping}
\end{figure}

Increasing hybridization with La-$5d$ states and decreasing hybridization with O-$2p$ states in LaNiO$_2$ will lead to electron transfer from the Ni-$3d$ states to La-$5d$ and O-$2p$ states, so the DFT+DMFT calculations give a Ni-$3d$ total occupancy of 8.593 that is greater than SrNiO$_2$ (8.479) by only 0.11 electrons, even though one more electron is residing in LaNiO$_2$. This suggests that the  Ni-$3d$ total occupancy is almost pinned from the undoped end to the extremely doped end, which has also been confirmed by our doping calculations. The overall change of the Ni-$3d$ total occupancy is only 0.08 in the doping range from $x=0$ to $x=0.5$ [see Fig.~\ref{fig:doping}(a)]. Comparing to SrNiO$_2$, the occupancy numbers among $3d$ orbitals in LaNiO$_2$ are slightly redistributed, with 0.162 (0.046) more electrons residing on $d_{x^2-y^2}$ ($d_{xz/yz}$) orbitals and 0.118 less electrons residing on $d_{3z^2-r^2}$ orbital, so still with substantial holes residing on these orbitals, as highlighted in the inset of Figs.~\ref{fig:band}(k). As a result, the probability distribution of the Ni-$3d$ multiplets shown in Table~\ref{tab:multi} is not too different between LaNiO$_2$ and SrNiO$_2$, so LaNiO$_2$ also shows multi-orbital nature and strong metallicity. The calculated weight of the spin-triplet configurations is 25.9\% for LaNiO$_2$, which is very close to the value, 24\%, given by the multiplet calculation used to simulate the x-ray absorption (XAS) and resonant inelastic x-ray scattering (RIXS) spectra of LaNiO$_2$ in~\cite{Hepting2020}. Their simulations show that substantial $3d^8$ spin-triplet configurations are responsible for the resonant pre-peak, $A^{\prime}$, of RIXS. The critical roles of spin-triplet configurations have also been pointed out in~\cite{Jiangmi:2019}. Different from~\cite{Karp:2020}, we obtain a small percentage of the $3d^{10}$ configuration (8.1\%) that rules out the possibility of the charge-transfer Mott insulator scenario in LaNiO$_2$. The percentages of other multiplets obtained here are also very close to the values given by~\cite{Hepting2020}, which validates our DFT+DMFT calculations.

Going from SrNiO$_2$ to LaNiO$_2$, the orbital differentiation is significantly enhanced, with stronger mass enhancement of 2.81 for $d_{x^2-y^2}$ orbital and weaker mass enhancement of about 1.2$\sim$1.3 for the other $3d$ orbitals in LaNiO$_2$. Fig.~\ref{fig:doping}(b) shows that a tendency of increasing correlation in $d_{x^2-y^2}$ orbital and decreasing correlation in other $3d$ orbitals, as decreasing hole-doping level $x$, has been captured by our doping calculations. 
As shown in the DFT+DMFT calculations by Lechermann~\cite{Lecherman:2020a,Lecherman:2020}, a large enough Hubbard $U$ ($\ge$ 10 eV) can open a Mott gap in the $d_{x^2-y^2}$ orbital. Our results strongly support their claim that the multi-orbital nature rules the normal state of NdNiO$_2$~\cite{Lecherman:2020a,Lecherman:2020}, but with strong metallicity in $d_{x^2-y^2}$ orbital found in our work. Fig.~\ref{fig:doping}(a) shows the evolution of the DFT+DMFT Fermi surface as increasing hole-doping. We find Lifshitz transitions of the Fermi surface from $x=0$ to $x=0.5$ and two sheets of Fermi surfaces around the optimal doping level, $x\approx0.2$. The results of occupancy numbers and mass enhancement of Ni $d_{x^2-y^2}$ and $d_{3z^2-r^2}$ orbitals as well as the evolution of Fermi surfaces as increasing hole-doping presented in this work are consistent with the previous DFT+DMFT study with $U=6$ eV, $J_H=0.95$ eV and fully localized DC scheme~\cite{Leonov:2020}.

\begin{figure}
\centering
\includegraphics[width=0.48\textwidth]{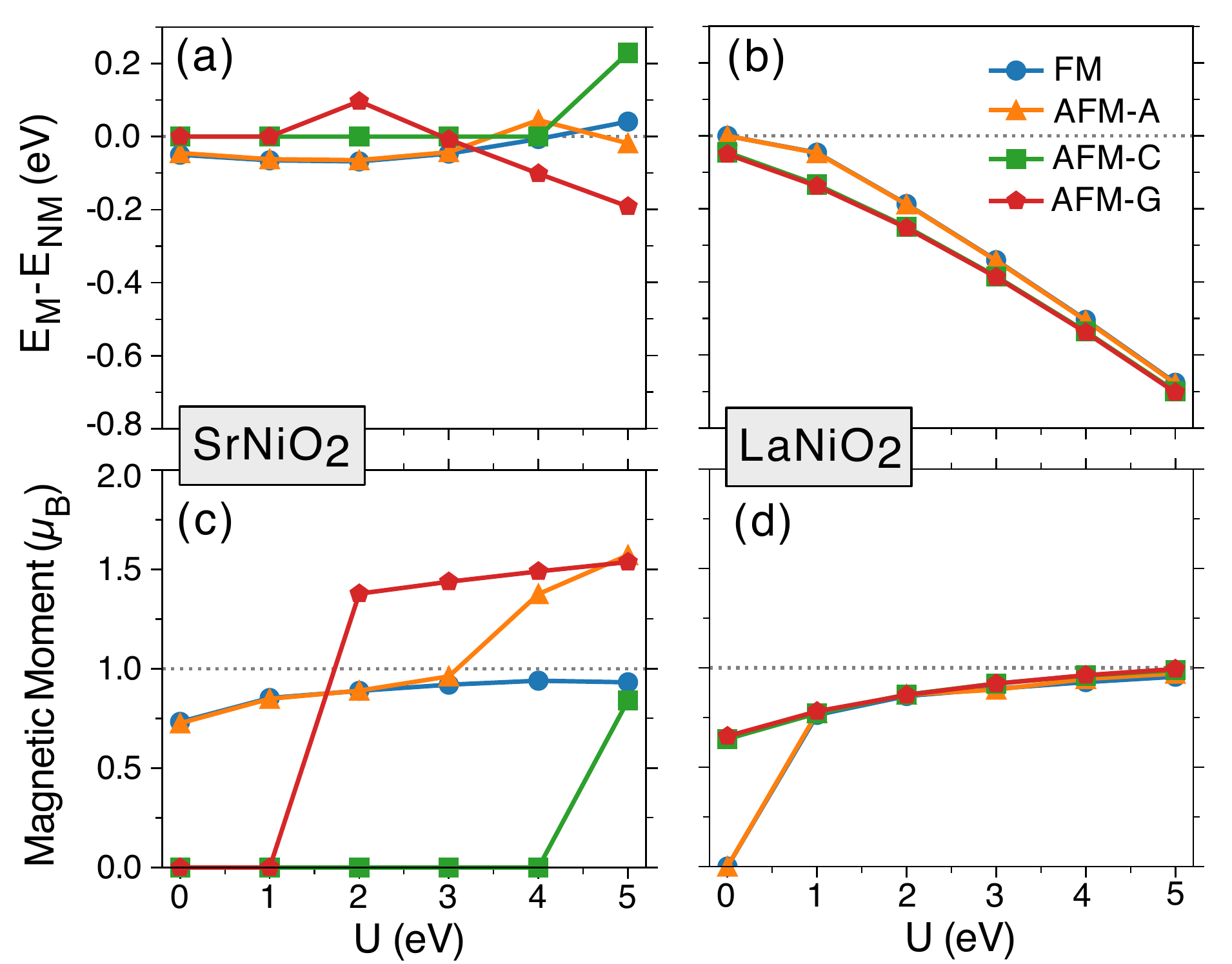}
\caption{(a),(b) The total energy difference (per the unit cell containing one Ni site) between the magnetic states and the non-magnetic state, 
$E_{\text{M}}-E_{\text{NM}}$, vs the Hubbard $U$
from the GGA+$U$ calculations. (c),(d) The ordered magnetic moments per one Ni site vs $U$. (a),(c) for SrNiO$_2$ and (b),(d) for LaNiO$_2$.}
\label{fig:mag}
\end{figure}

\textit{Magnetism}---One of the main arguments to question the cuprate-like superconducting nature in Nd$_{1-x}$Sr$_x$NiO$_2$ is that the super-exchange coupling $J_{\text{ex}}$ is too small due to the very large $\Delta_{dp}$~\cite{Jiangmi:2019}. Here, like CaCuO$_2$~\cite{Botana:2020,Karp:2020}, $\Delta_{dp}$ is significantly decreased in SrNiO$_2$, which may stabilize long-range AFM orders or induce stronger AFM fluctuations through an enhanced $J_{\text{ex}}$. We  check this by  comparing the ground state energy of possible magnetic configurations in SrNiO$_2$ to that in LaNiO$_2$ on the static mean-field (GGA+$U$) level. Figs.~\ref{fig:mag}(a) and \ref{fig:mag}(b) show the energy difference per one Ni site between the magnetic and non-magnetic (NM) states, $\delta E=E_{\text{M}}-E_{\text{NM}}$, as functions of Hubbard $U$. For SrNiO$_2$, the lowest energy state is FM and AFM-A when $U\le3$ eV and AFM-G when $U>3$ eV. For LaNiO$_2$, AFM-G is the lowest energy state, even at the GGA level ($U=0$)~\cite{Ryee:2020}. FM, AFM-A and AFM-C are also lower in energy than the non-magnetic state when $U>0$. $\delta E$ of LaNiO$_2$ becomes much larger than that of SrNiO$_2$ when $U>0$, which suggests that it is much easier for LaNiO$_2$ to stabilize magnetic orders than SrNiO$_2$, hence the physics of super-exchange is not operational in these materials, while they are essential for the cuprates. 

Although the static mean-field calculations can give stable long-range magnetic orders with large ordered magnetic moments [see Figs.~\ref{fig:mag}(c) and~\ref{fig:mag}(d)] in both of LaNiO$_2$ and SrNiO$_2$, in reality, no signature of such orders have been observed in LaNiO$_2$ or NdNiO$_2$ down to the very low temperature range the available experiments have approached so far~\cite{Hayward:1999,Hayward:2003}. This situation is similar to that in iron-based superconductors, where the static mean-field calculations usually predict stable magnetic orders with large ordered moments while the ordered moments are found to be significantly suppressed in experiments, as discussed widely in literatures such as~\cite{Ishibashi:2008,Huang:2008,Ishida:2009,Qureshi:2010,Borisenko:2010,yin:2011,Hansmann:2010,Toschi:2012}.


\textit{Conclusion}---To summarize, we have provided a new perspective to describe the normal state electronic structure of Nd$_{1-x}$Sr$_x$NiO$_2$ starting using SrNiO$_2$ as a reference. SrNiO$_2$ is found to be a moderately correlated multi-orbital Hund-like metal. Going from SrNiO$_2$ to LaNiO$_2$, the  Ni-$3d$ total occupancy is insensitive to decreasing hole-doping as the changes in carrier density are compensated by decreased hybridization with the occupied O-$2p$ states and increased hybridization with the unoccupied La-$5d$ states. Thus, as in SrNiO$_2$, same amount of high-spin configurations, same order of charge fluctuation and similar multi-orbital nature but with enhanced orbital differentiation, are found in LaNiO$_2$. Our results are in good agreement with the existing RIXS measurements.

From this perspective, our study excludes the cuprate-like scenario for this new Ni-based superconductor and suggests that Nd$_{1-x}$Sr$_x$NiO$_2$ is a new unconventional superconductor that shares many similarities with Hund's superconductors~\cite{Hoshino:2015,Werner:2020,Tsunghan:2018}, such as their metallicity, the importance of Hund's coupling and high-spin configurations, tendency towards orbital differentiation with the $d_{x^2-y^2}$ as the most correlated orbital, and the absence of magnetism in regimes which are ordered according to static mean-field theories, but with a valence close to $3d^{8.5}-3d^9$ and with low-energy $e_g$ orbitals instead of the Fe $d^{5.5}-d^{6.5}$ valence with low-energy $t_{2g}$ orbitals~\cite{Kamihara:2008,stewart:2011,Qimiao:2016}. Most interestingly, the ratio between the largest superconducting gap $2\Delta_{max}$ and $T_{c}$ (determined as the mid-point of the superconducting transition in resistivity) is evaluated to be about 7 from the most recent experimental data of Nd$_{1-x}$Sr$_x$NiO$_2$~\cite{Haihu:2020}, which is very close to the value for the iron-based superconductors that are proposed as prototypical Hund's superconductors~\cite{Miao2018,Tsunghan:2018}. 

Finally, we notice that as we place Nd$_{1-x}$Sr$_x$NiO$_2$ far from a Mott transition, the  insulating behavior observed at low doping and temperatures, has to be attributed  to the impurity induced Anderson localization, which should get weaker as sample quality improves.

\textit{Note added.}---Under the preparation of this manuscript, we learned about the related parameter-free $GW$+EDMFT work~\cite{Francesco:2020} that also finds multi-orbital nature and the pinning of the Ni-$3d$ total occupancy in the small doping regime of Nd$_{1-x}$Sr$_x$NiO$_2$, which validates our simple DFT+DMFT calculations.

\textit{Acknowledgment.}---We thank Kristjan Haule and Sangkook Choi for very helpful discussions. Y. W., C. K. and G. K. were supported by the U.S. Department of Energy, Office of Science, Basic Energy Sciences as a part of the Computational Materials Science Program through the Center for Computational Design of Functional Strongly Correlated Materials and Theoretical Spectroscopy. H. M. was supported by the Laboratory Directed Research and Development (LDRD) of ORNL, under project No. 10018.

\bibliography{main}

\end{document}


\title{Supplementary Materials for ``Hund's metal physics: from SrNiO$_2$ to NdNiO$_2$''}
\author{Y. Wang}     
\affiliation{Department of Condensed Matter Physics and Materials Science, Brookhaven National Laboratory, Upton, New York 11973, USA} 

\author{C.-J. Kang}
\affiliation{Department of Physics and Astronomy, Rutgers University, Piscataway, New Jersey 08856, USA}

\author{H. Miao}
\affiliation{Material Science and Technology Division, Oak Ridge National Laboratory, Oak Ridge, Tennessee 37830, USA} 

\author{G. Kotliar}
\affiliation{Department of Condensed Matter Physics and Materials Science, Brookhaven National Laboratory, Upton, New York 11973, USA} 
\affiliation{Department of Physics and Astronomy, Rutgers University, Piscataway, New Jersey 08856, USA}

\date{\today}

\begin{abstract}
\end{abstract}

\maketitle

\section{Computational Details}
\begin{figure}
\centering
\includegraphics[width=0.49\textwidth]{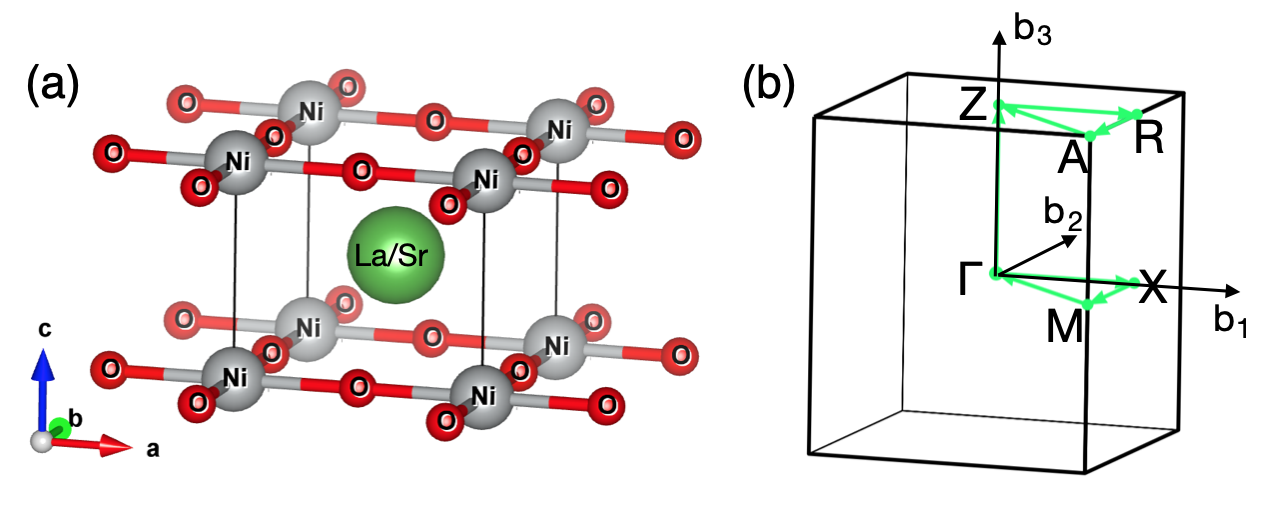}
\caption{(a) The crystal structure of LaNiO$_2$ or SrNiO$_2$. (b) The Brillouin zone and high-symmetry $K$-path.}
\label{fig:struct}
\end{figure}

\begin{figure*}
\centering
\includegraphics[width=0.9\textwidth]{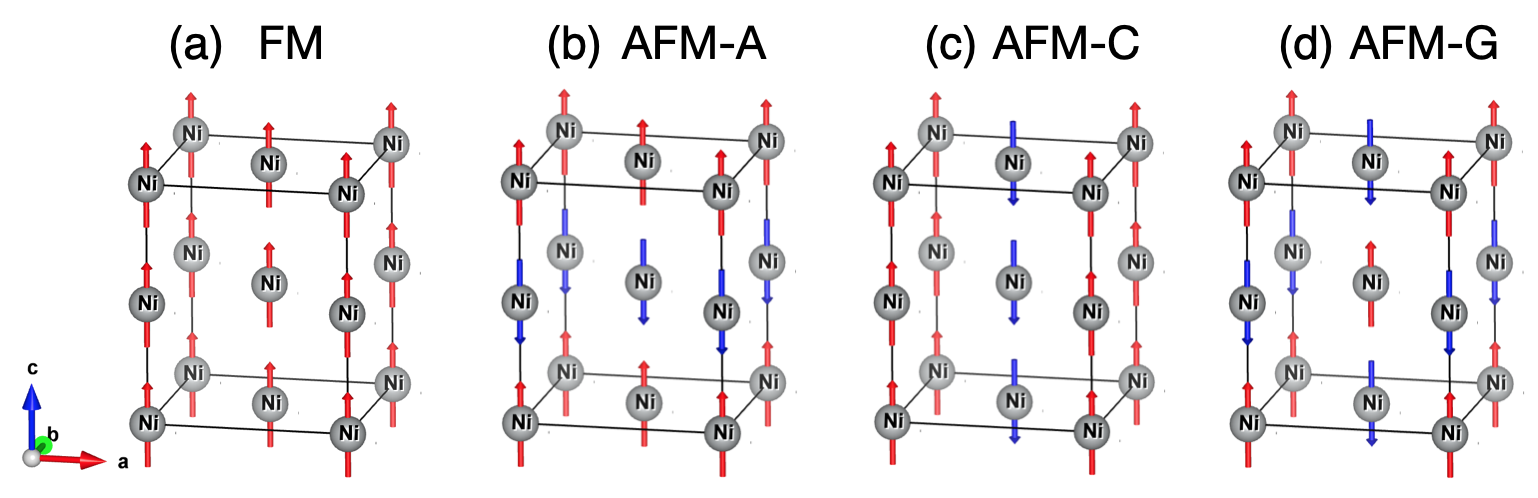}
\caption{Magnetic configurations with a $\sqrt{2}\times\sqrt{2}\times2$ supercell. Arrows indicate spin directions on the Ni sites.
(a) ferromagnetic (FM), (b) ferromagnetic in plane and antiferromagnetic out of plane (AFM-A), (c) antiferromagnetic in plane and ferromagnetic out of plane (AFM-C) and (d) antiferromagnetic in plane and antiferromagnetic out of plane (AFM-G).}
\label{fig:magconfig}
\end{figure*}

Fig.~\ref{fig:struct}(a) shows the crystal structure of La(Sr)NiO$_2$ generated by the VESTA software~\cite{Momma:2008}. Fig.~\ref{fig:struct}(b) shows the corresponding Brillouin zone (BZ) and the high-symmetry $K$-path used in the band structure plots. The spin-polarized calculations are performed on the static mean-field level using the DFT+$U$ method implemented in the VASP package~\cite{kresse:1996} with projector augmented-wave (PAW) pseudopotential~\cite{blochl:1994,kresse:1999} and Perdew-Burke-Ernzerhof parametrization of the generalized gradient approximation (GGA-PBE) exchange-correlation functionals~\cite{perdew:1996}. The energy cutoff of the plane-wave basis is set to be 600 eV, and a $\Gamma$-centered $20 \times 20\times 20$ $K$-point grid is used. Fig.~\ref{fig:magconfig} shows the magnetic configurations considered in the calculations.

\section{Fermi Surface}
\begin{figure*}
\centering
\includegraphics[width=0.98\textwidth]{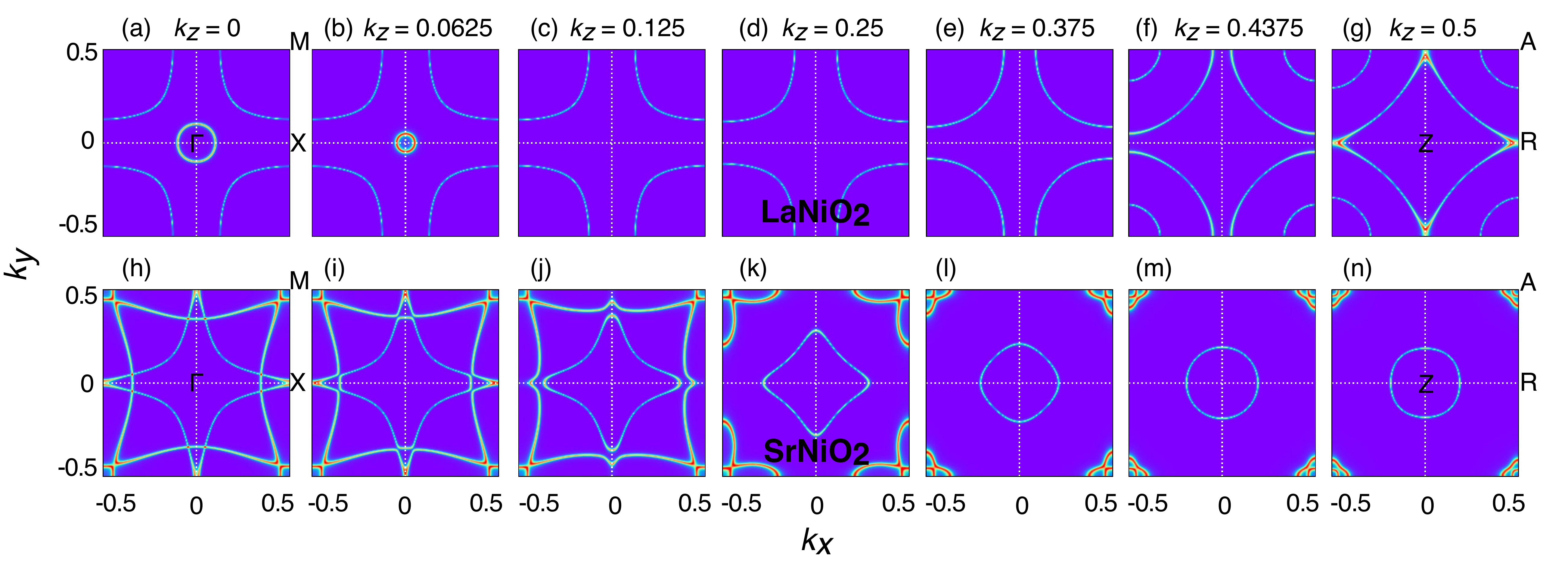}
\caption{The DFT+DMFT Fermi surfaces at $k_z=0, 0.0625, 0.125, 0.25, 0.375, 0.4375, 0.5$. (a)-(g) for LaNiO2 and (h)-(n) for SrNiO2.}
\label{fig:fs}
\end{figure*}

Fig.~\ref{fig:fs} shows the DFT+DMFT Fermi surfaces (FSs) cuts at $k_z=0, 0.0625, 0.125, 0.25, 0.375, 0.4375, 0.5$ for LaNiO$_2$ (upper panel) and SrNiO$_2$ (lower panel). The FSs of LaNiO$_2$ contain three sheets~\cite{Leonov:2020,Karp:2020}: (1) an electron-like pocket around $\Gamma$-point with mainly Ni-$d_{3z^2-r^2}$ and La-$5d$ characters; (2) another electron-like pocket around $A$-point with mainly Ni-$d_{xz/yz}$ and interstitial La-$s$ characters; (3) a hole-like FS centered along the $A$-$M$ BZ edge with Ni-$d_{x^2-y^2}$ character, which is just at the von Hove singularity point [see Fig.~\ref{fig:fs}(g) at $k_z=0.5$], indicating Lifshitz transition upon hole-doping. The FSs of SrNiO$_2$ also contain three sheets: (1) an electron-like FS around $\Gamma$-point with mainly Ni-$d_{x^2-y^2}$ character; (2) a hole-like FS centered along $A$-$M$ BZ edge with mainly Ni-$d_{xz/yz}$ character; (3) another hole-like FS centered along $A$-$M$ BZ edge with mainly Ni-$d_{xz/yz}$ and -$d_{3z^2-r^2}$ characters.

\bibliography{suppl}